# Energetic, spatial and momentum character of the electronic structure at a buried interface: the two-dimensional electron gas between two metal oxides

S. Nemšák[1,2,#], G. Conti[1,2], A.X. Gray[1,2,3,@], G.K. Palsson[1,2,&], C. Conlon[1,2], D.Eiteneer[1,2], A. Keqi[1,2], A. Rattanachata[1,2], A.Y. Saw[1,2], A. Bostwick[2], L. Moreschini[2], E. Rotenberg[2], V.N. Strocov[4], M. Kobayashi[4], T. Schmitt[4], W.Stolte[2,♣], S. Ueda[5], K. Kobayashi[5], A. Gloskovskii[6], W. Drube [6], C. Jackson[7,♥], P. Moetakef[7,♦], A. Janotti[7], L. Bjaalie[7], B. Himmetoglu[7], C. G. Van de Walle[7], S. Borek[8], J. Minar[8,9], J. Braun[8], H. Ebert[8], L. Plucinski[10], J.B. Kortright[2], C.M. Schneider[10], L. Balents[11], F.M.F. de Groot[12], S. Stemmer[6], and C. S. Fadley[1,2]

[1]Department of Physics, University of California, 1 Shields Ave, Davis, CA-95616, USA

[2]Materials Sciences Division, Lawrence Berkeley National Laboratory, 1 Cyclotron Rd, Berkeley, CA-94720, USA

[3]Stanford Institute for Materials and Energy Science, Stanford University and SLAC National Accelerator Laboratory, 2575 Sand Hill Road, Menlo Park, CA 94029, USA

[4]Swiss Light Source, Paul Scherrer Institute, Villigen, Switzerland

[5] NIMS Beamline at SPring-8, National Institute for Materials Science, Hyogo 679-5148, Japan

[6]Deutsches Elektronen-Synchrotron DESY, 22607 Hamburg, Germany

[7]Materials Department, University of California, Santa Barbara, CA-93106, USA

[8]Physical Chemistry Institute, Ludwig-Maximilian University, Munich, Germany

[9]New Technologies-Research Center, University of West Bohemia, 306 14 Plzen, Czech Republic

[10] Peter-Grünberg-Institut PGI-6, Forschungszentrum Jülich, 52425 Jülich, Germany

[11] Department of Physics, University of California, Santa Barbara, CA-93106, USA

[12] Inorganic Chemistry and Catalysis, Utrecht University, 3584 CG Utrecht, Netherlands

[#]Present address: Peter-Grünberg-Institut PGI-6, Forschungszentrum Jülich, 52425 Jülich, Germany

[@]Present address: Department of Physics, Temple University, Philadelphia, PA 19122, USA

[♣]Present address: National Security Technologies, Livermore, CA 94551 USA

[&]Present address: Department of Physics, Uppsala University, Uppsala, SE-751 20 Sweden

[♥]Present address: HRL Laboratories, LLC, Malibu, CA 90265 USA

[♦]Present address: Department of Chemistry, University of Maryland, College Park, MD 20742, USA

**ABSTRACT**

The interfaces between two condensed phases often exhibit emergent physical properties that can lead to new physics and novel device applications, and are the subject of intense study in many disciplines. We here apply novel experimental and theoretical techniques to the characterization of one such interesting interface system: the two-dimensional electron gas (2DEG) formed in multilayers consisting of $SrTiO_3$ (STO) and $GdTiO_3$ (GTO). This system has been the subject of multiple studies recently and shown to exhibit very high carrier charge densities and ferromagnetic effects, among other intriguing properties. We have studied a 2DEG-forming multilayer of the form [6 unit cells STO/3 unit cells of GTO]$_{20}$ using a unique array of photoemission techniques including soft and hard x-ray excitation, soft x-ray angle-resolved photoemission, core-level spectroscopy, resonant excitation, and standing-wave effects, as well as theoretical calculations of the electronic structure at several levels and of the actual photoemission process. Standing-wave measurements below and above a strong resonance have been exploited as a powerful method for studying the 2DEG depth distribution. We have thus characterized the spatial and momentum properties of this 2DEG with unprecedented detail, determining via depth-distribution measurements that it is spread throughout the 6 u.c. layer of STO, and measuring the momentum dispersion of its states. The experimental results are supported in several ways by theory, leading to a much more complete picture of the nature of this 2DEG, and suggesting that oxygen vacancies are not the origin of it. Similar multi-technique photoemission studies of such states at buried interfaces, combined with comparable theory, will be a very fruitful future approach for exploring and modifying the fascinating world of buried-interface physics and chemistry.

**I. INTRODUCTION**

There is presently high interest in bilayer and multilayer structures involving metal oxides, due to the novel electronic states that can develop at the interfaces between the different constituents, which can often be markedly different from those states in the native materials[1,2,3]. A by now classic example of this is the 2-dimensional electron gas (2DEG) at the $SrTiO_3/LaAlO_3$ heterointerface[1], which was revealed to be conducting, and also to exhibit magnetic properties[2,4] as well as superconductivity[2,5], in spite of both constituents being non-magnetic insulators. The electronic structure of this 2DEG has also been studied by resonant soft x-ray angle-resolved photoemission (ARPES)[6], a technique that we expand below to include resonant standing-wave excitation. Many other oxide heterostructure systems have been studied to date, and, beyond exhibiting much new physics, have also shown promise for future logic or storage devices[3,7]. It is thus not an overstatement to repeat that "The interface is the device"[8,9], in many important applications areas. These include for example interfaces involving a solid oxide and a liquid electrolyte that have been studied in gated devices[10], but which are also crucial in energy production, electrochemistry, corrosion, and environmental science[11,12,13]; such solid/liquid interfaces have been studied for many years but are still not fully understood.  The quantitative characterization of solid/solid or solid/liquid interfaces is thus a challenge in many areas of science and technology. We will focus here on an illustrative specific example of a 2DEG formed at an oxide interface, studied with novel photoemission techniques and theoretical modeling, but with general implications for a wide variety of future interface studies.

The synthesis of any interface, but particularly the oxide interfaces, is critical, with control of epitaxy, strain, interface sharpness, and oxygen concentration all having a strong influence on the existence and properties of the 2DEG[3]. Several explanations of the 2DEG origin at such complex oxide interfaces and their free surfaces in vacuum have been given in recent years, ranging from oxygen deficiency at the surface/interface to charge transfer among layers to avoid the polar catastrophe[1,14]. A complete understanding of the nature of such 2DEGs is needed in order to further exploit and control their properties.  We have thus applied a unique combination of photoemission measurements, including soft and hard x-ray energies, standing-wave and resonant excitation, and angle-resolved photoemission (ARPES) to determine the detailed energy-, momentum-, and depth-resolved nature of such a 2DEG at a buried interface between the band insulator $SrTiO_3$ (STO) and the ferromagnetic Mott insulator $GdTiO_3$ (GTO).

The STO/GTO system has been studied extensively by Stemmer and co-workers[15,16,17,18,19,20,21,22,23,24,25,26], and the 2DEG here has been found to exhibit extremely high carrier densities and ferromagnetic effects, with both electrostatic and doping modulation being observed to change the carrier properties.  The proposed explanation for the 2DEG creation can be found in a simple charge transfer picture[7,14,16].  In particular, The GdO and $TiO_2$ planes in $GdTiO_3$ carry +1 and −1 formal ionic charges, respectively. Each GdO layer in $GdTiO_3$ can be considered to donate one-half an electron to the $TiO_2$ planes above and below it, including the interfacial $TiO_2$ plane, in which the half-electron forms a mobile 2DEG state.  Looking at the interface from this perspective of GTO, every terminating unit cell of GTO is losing half of an electron. These donated d- electrons of titanium occupy conduction-band states in the STO and act as free electrons.   The accuracy of this picture is confirmed by our experimental standing-wave results and first-principles calculations, as described in Sec. III.D.  The presence of oxygen vacancies is not needed for the explanation of this 2DEG formation, although such vacancies at interfaces are believed to play a role in the LAO/STO and other systems[6,27], and we comment on their relevance to GTO/STO in our conclusions.  Beyond this, the strong role of structural distortions and symmetry breaking near the interface have been pointed out[22,25,28,29].

In this paper, we have applied a broad variety of photoelectron techniques, together with cutting-edge density functional theory (DFT) calculations, to determine for the first time the energy-resolved, momentum-resolved, and depth-resolved character of such a 2DEG, with each technique providing a complementary piece of information in this determination, and theory supporting and clarifying our conclusions in several ways.

In order to probe more deeply into the buried interface, photoemission with both soft x-ray (~460-1200 eV) and hard x-ray excitation (~6000 eV) was used. At these energies the inelastic mean free paths (IMFPs) of the photoelectrons that determine the information depth range in STO(GTO) from 11 Å(10 Å) at 460 eV to 22 Å(20 Å) at 1200 eV to 83 Å(72 Å) at 6000 eV (as determined from the TPP-2M formula using the SESSA program[30]), thus permitting us to penetrate into the first buried interface of our multilayer, which is about 20 Å below the surface, as shown in Fig. 1(a). Such considerations have also led to increased interest in carrying out angle-resolved photoemission (ARPES) in the soft x-ray regime[31], and we have done that in this study.

In another set of measurements, in order to enhance photoemission of states with Ti 3d character, the well-known resonant photoemission at the Ti 2p-to-Ti 3d absorption edge at about 465 eV was used.

For precise depth resolution of the electronic structure, the recently developed standing-wave approach, including momentum resolution via standing-wave angle-resolved photoemission (SWARPES) was exploited[32,33]. In this technique, the incidence angle $\theta_x$ is tuned to the first-order Bragg condition of the multilayer as determined from $\lambda_x = 2d_{ML}sin\theta_{Bragg}$, where $\lambda_x$ is the x-ray wavelength, $d_{ML}$ is the bilayer thickness in the multilayer mirror, and $\theta_{Bragg}$ is the Bragg angle, so that a strong standing wave (SW) is created. The incidence angle is then scanned through that angle, thus also scanning the SW by one-half of its wavelength vertically through the sample, as indicated in Fig. 1(a). Such scans generate rocking curves (RCs) of core and valence photoelectron intensities that encompass all the elements in the sample. Additional interference effects due to reflection of the x-rays from the top and bottom of the multilayer sample generate fine structure termed Kiessig fringes[32,33]. If the multilayer sample has a total thickness of $D_{ML} = Nd_{ML}$, where $N$ is the number of bilayer repeats in the multilayer mirror, the interference maxima will appear for $m\lambda_x = 2D_{ML}sin\theta_{Kiessig,m}$, where m is some high order of interference and $\theta_{Kiessig,m}$ is the incidence angle corresponding to this order. Additional resonant effects on the x-ray optical constants <u>below</u> and <u>above</u> the Gd $3d_{5/2}(M_5)$ absorption edge were also used to move the scanning depth of the SW in a controlled way and to increase the amplitude of the SW. Moving the standing-wave position in this way over a resonance was first demonstrated by Bedzyk and Materlik[34] for Bragg reflections from single-crystal planes and with x-ray fluorescence detection; this work is the first time to our knowledge that this technique has been used with reflection from a multilayer heterostructure and with photoelectron detection. The standing-wave data is compared to simulations using the Yang X-Ray Optics (YXRO) software package[35] which includes detailed x-ray optical and photoemission intensity modeling, and implicitly both Bragg and Kiessig interference effects. The expected depth resolution of the SW technique with fitting of experimental rocking curves to x-ray optical theory while varying geometric parameters is expected to be $\leq 1/10^{th}$ of the SW period, which is comparable to the size of 1 u.c. of STO (3.905 Å).

The experimental results for electronic structure are then compared to state-of-the-art hybrid functional calculations[7, 36,37] and one-step photoemission calculations of ARPES in the relativistic spin-polarized KKR approach that includes the surface and matrix-element effects[38]. We also discuss core-level binding energy shifts and their theoretical interpretation[39], e.g. using an Anderson Impurity Model with final-state screening[40].

## II. EXPERIMENTAL METHODS

Reference samples of STO (30 nm thick) and GTO (26 nm thick) of thicknesses much greater than the photoelectron inelastic mean free paths (IMFPs), as well as the primary epitaxial multilayer STO/GTO sample were prepared using a hybrid molecular beam epitaxy approach with both gas and solid sources[15,20]. The thick reference samples are thus representative of "bulk" material since the photoemission measurement does not reach the bottom of the layer in sensitivity. The multilayer sample was grown on a (001) surface of $(LaAlO_3)_{0.3}(Sr_2AlTaO_6)_{0.7}$ (LSAT) single crystals with Ta-backing layers in a GEN930 oxide MBE system, and a chamber background pressure of $10^{-9}$ torr. The geometry of the sample and the photoemission experiment are shown in Figure 1(a). Twenty bi-layers, each consisting of 6 unit cells (u.c.) of STO (nominally 23.7 Å) and 3 u.c. of GTO (nominally 11.8 Å) were deposited on the LSAT substrate and capped by a final STO layer (5 u.c. or 19.8 Å thick). The period of the multilayer is thus 35.5 Å, which will also be the period of the SW perpendicular to the surface of the sample[32,33]. The STO and GTO layers were coherently and compressively strained to the underlying LSAT by about 1%. To clarify the precise sample structure, Fig. 1(b) shows the internal stacking of $TiO_2$, SrO, and GdO layers, which make it clear that each interface contains a shared $TiO_2$ layer. Scanning transmission electron microscopy measurements with high angular aperture dark field imaging confirm well-defined and sharp interfaces, with very little inter-diffusion, as demonstrated for a similar sample in Ref.15. The results of x-ray optical simulations of core-level RCs also confirm the high quality of the sample and its interfaces, as shown below.

Photoemission measurements were carried out at several facilities, permitting measurements with both soft and hard x-rays excitation. In all cases, the radiation was p-polarized, with different fixed angles $\theta_{xe}$ between x-ray incidence and electron exit, as indicated in Fig. 1(a) and specified below. The soft x-ray data was obtained at Beamline 7.0.1 of the Advanced Light Source (Ti $2p_{3/2}$ or $L_3$ resonant at ~460 eV and non-resonant at 833 eV, with an overall energy resolution of 0.15 eV and $\theta_{xe} = 60°$) and the ADRESS Beamline at the Swiss Light Source (Gd $3d_{5/2}$ or $M_5$ resonant at 1182 or 1187 eV, resolution of 0.20 eV and $\theta_{xe} = 70°$)[41], and the hard x-ray data was obtained at the P09 Beamline of Petra III (~5000 eV, resolution of 0.30 eV and $\theta_{xe} = 88°$), Beamline BL15XU of SPring-8 (~6000 eV, resolution of 0.30 eV and $\theta_{xe} = 88°$) and Beamline 9.3.1 of the Advanced Light Source (~4000 eV, resolution of 0.60 eV and $\theta_{xe} = 88°$ ). Frequent calibration of the Fermi level ($E_F$) was made with gold samples, and all experimental binding energies are presented with this reference. Each of these experimental stations is equipped with a hemispherical electrostatic spectrometer manufactured by Scienta or Specs and sample manipulators that can scan the polar angle $\theta$ and the azimuthal angle $\phi$ in Fig. 1(a), and, in the case of the soft x-ray measurements, also the "tilt" angle $\beta$ that is critical to obtaining ARPES results versus $k_x$ and $k_y$.

### III. EXPERIMENTAL AND THEORETICAL RESULTS AND DISCUSSION

### III.A. Valence-band densities-of-states and core-level spectra:

In Fig. 2, we show room-temperature soft- and hard- x-ray valence-band photoemission at 833 eV (Fig. 2(a)) and between 4000 and 5950 eV (Fig. 2(c), for bulk STO, bulk GTO, and the multilayer). At this temperature, such spectra should represent matrix element-weighted densities of states (MEWDOS). We can here identify, particularly in the soft x-ray spectrum of 2(a) and its blowup in 2(b), five features 1, 1′, 2, 3, 4 and 5, with these being slightly different in position and much different in weight with hard x-ray excitation due to matrix-element (or equivalently cross- section) effects. The feature labels in the hard x-ray spectra of Fig. 2(c) are color-coded to indicate the dominant origin, either from STO (green) or GTO (blue). The blowup in Fig. 2(d) again shows for the multilayer spectrum (red curve) features reminiscent of 1 and 1′ in Fig. 2(b), with the spectrum of GTO being more similar to feature 1′ in the soft x-ray data, and STO not showing significant intensity over this near-Fermi-energy region, as expected for STO with a bandgap of ca. 3.4 eV. Figs. 2(e) and 2(f) show soft x-ray core-level spectra for the two elements characteristic of each member of the bilayer, Ti 2p and Sr 3d, respectively. Ti2p clearly shows evidence of two valence states, $Ti^{4+}$ associated with the top layer STO and all other STO layers, and $Ti^{3+}$ associated primarily with the GTO layers, or possibly also with the presence of $Ti^{3+}$ near STO/GTO interfaces. Rocking curve data below conclusively show that $Ti^{4+}$ is associated with STO, and $Ti^{3+}$ is associated with GTO. Although not as pronounced as for Ti, peak fitting of the Sr 3d spectra in Fig. 2(f) indicate a higher-binding-energy (HBE) state that is about 1/3 as intense as the main $Sr^{2+}$ low-binding-energy (LBE) peak and with a separation of about 0.7-0.8 eV. Such HBE and LBE components for Sr 3d have in fact been observed before in an XPS study of $SrTiO_3$[42], with the HBE peak interpreted as being due to surface Sr species bound in more electronegative environments to F and O. Alternatively, the possibility of different Sr electronic configurations such as $Sr^{2+}$ and $Sr^{1+}$ associated with final-state screening of the Sr 3d core hole have been considered[39]. We comment further below on the behavior of these core-level components using the additional information from the standing-wave scans through the sample.

Figure 3(a) shows the calculated ferromagnetic band structure for bulk GTO, which has a 20-atom unit cell with the $GdFeO_3$ crystal structure; octahedral distortions are fully taken into account [Fig. 3(b)]. Lower and upper Hubbard bands (LHB and UHB) are indicated in the band structure. These calculations were done using DFT with a hybrid exchange-correlation functional capable of addressing the strong electron-electron interactions that govern the electronic properties of GTO[36,37]. Conventional exchange-correlation functionals, such as the local density approximation (LDA) or generalized gradient approximation (GGA), yield an incorrect description of the electronic structure of Mott insulators; in this case, they predict GTO to be a metal. These functionals suffer from spurious self-interaction and tend towards excessive delocalization of the electronic density. Hybrid functionals, on the other hand, provide partial cancellation of self-interaction, and the HSE functional has been shown to give an accurate description of the electronic and structural properties of a wide range of materials[43], including perovskite oxides[44].

The calculations were done with the VASP code[45,46], using the default 25% Hartree-Fock mixing parameter and a 0.2Å$^{-1}$ range separation parameter. Projector augmented wave potentials (PAW)[47] were used to describe the interaction between the valence electrons and the ionic cores, with the Gd 4f electrons included in the valence bands for bulk GTO and treated as core levels for the superlattice. All calculations were carried out with spin polarization, which is required to correctly describe the electronic properties of GTO. The bulk calculation (Fig. 3(a)) was done using a 500 eV plane-wave energy cutoff and a 4x4x2 Monkhorst-Pack k-point grid, and the superlattice calculation (Fig. 9) was done with a 400 eV cutoff, and a 4x4x1 k-point grid.

Fig. 3(c) shows total and projected densities of states from all of the elements in bulk GTO, and it is clear that the LHB is mainly Ti 3d, but with strong mixing of O 2p, and to some degree also Gd 5d. In Fig. 3(d), the total densities of states for GTO and STO are shown, aligned using the calculated band offset of 2.6 eV that is in good agreement with the 2.9 ± 0.3 eV found from hard x-ray photoemission measurements[21]. From Figs. 2 and 3, we can thus tentatively identify peak 1' in experiment as being due to the LHB, and peaks 2, 3, and 4 in Figs. 2(a) and 2(c) to the peaks in the combined densities of states of STO and GTO in Fig. 3(d) denoted a, b, and c, respectively. Peak 5 in experiment is assigned to Gd 4f states. The experimental results also make it clear that the same LHB bands are observed both in bulk GTO and the multilayer (Figs. 2(b) and (d)). However, peak 1, which is centered about 350 meV below the Fermi level and marked with a dashed vertical guideline in Fig. 2(b), is not observed at all in the calculated spectra for bulk GTO; we show below that it is in fact a signature of the 2DEG.

**III.B. Resonant angle-resolved photoemission:**

We now turn to resonant photoemission and ARPES to explore more of the properties of the two states represented by features 1 and 1', including their **k**-space distributions. The methodology here is thus like that applied to the LAO/STO system by Berner et al.[6] Fig. 4(a) displays a series of angle-integrated MEWDOS-like valence-band (VB) spectra measured at liquid $N_2$ temperature as the photon energy sweeps through the Ti 2p resonances, with the peaks labelled pre = pre-edge, and a-d. Peaks a-d occur associated with the $2p_{1/2}$ ( $L_2$ ) and $2p_{3/2}$ ( $L_3$ ) absorption edges, with each of these being further split into their $e_g$ (b and d) and $t_{2g}$ (a and c) components; these features are well-known from prior x-ray absorption studies of Ti oxides[48]. The strong enhancement of features 1 and 1' is also obvious in Fig. 4(a), near the energies of peaks b and d. The maximum enhancement is found just below peak d, the Ti $L_2$ $e_g$ resonance, at a photon energy of 465.2 eV. Fig. 4(b) shows the resonantly enhanced MEWDOS VB spectrum, with both features 1 and 1' clearly visible. To explore the energy dependence of resonant- and non-resonant ARPES, we also define the five energy intervals A-E in Fig. 4(b) that span features 1 and 1', with A being strongly associated with feature 1.

We first consider three ARPES patterns over region A in Figs. 4(c)(i)-(iii), as measured off- and on- resonance. Measurements performed at a higher non-resonant excitation energy of 833 eV that will penetrate at least two interfaces (4(c)(i)), show a weak rotated square-grid pattern. The higher-energy results show over 20 Brillouin zones (BZs) due to the longer final-state **k**-vector and the higher angular acceptance in $\theta$, ($\pm 20°$) of the particular spectrometer used, with the associated mechanical $\beta$ scan being chosen to match this angular acceptance (cf. Fig. 1). On the other hand, for an energy off resonance at a lower photon energy 470 eV that is well above the Ti 2p resonances (4(c)(ii)), scarcely any modulation in the intensity is observable. However, going to the optimum resonance energy of 465.2 eV reveals an intense and sharp square-grid pattern (4(c)(iii)), which for this lower energy and the same spectrometer samples over 10 BZs. Using resonant excitation is thus essential for carrying out ARPES on these weak features at such low electron kinetic energies due to the rapidly decreasing mean free paths for both inelastic and elastic scattering, with elastic scattering further acting to smear features out in momentum[31].

As noted above, at these soft x-ray photon energies, we sample several BZs, and an analysis of the emission process in reciprocal space below reveals that the intensity maxima in Fig. 4(c) are located at $\Gamma$ points in each BZ, with sleeves of higher intensity running between those maxima through X directions, as labelled in Fig. 4(c)(iii). The difference in the intensity between different gamma points is caused by a different position in the BZ along the perpendicular direction of **k** caused by the curvature of the sphere in **k**-space that is being sampled at these energies, as discussed in more detail below. It is remarkable that the sharp square momentum distribution pattern in Fig. 4(c) is not only observed in the lowest BE interval of A, but also continues to appear for deeper states located as low as BE =-1.6 eV, although with reduced sharpness and intensity as binding energy increases.

An additional comment here concerns the periodicity of the patterns seen in Figs. 4(c)(i)-(vi), for which the $\Gamma$ and X labels are found to be those of the simple five-atom unit perovskite unit cell of GTO, rather than the larger unit cell that allows for octahedral tilts shown in Fig. 3(b). Such observations of effectively reduced spatial periodicity in ARPES data, as seen e.g. in the reduced importance of folded bands, have been discussed recently in an Fe-based superconductor[49] and a half-metallic ferromagnet[50]. We suggest that this is due at least partially to the final state damping with distance in ARPES due to inelastic scattering, which effectively reduces the spatial extent of sampling the long-range symmetry of the lattice, especially when the distortions from the simple unit cell are small. An additional cause is no doubt phonon effects, which are more important in the several-hundred eV or higher range, and tend to smear the experimental results in the BZ, thus again effectively shrinking the periodicity in experiment. Such phonon effects and their theoretical modeling are discussed in detail elsewhere[31,51].

Feature 1 and feature 1' exhibit similar **k**-space distributions following the periodicidy of the simple perovskite Brillouin zone, in spite of the fact that they have a different fundamental origin. This is consistent with the calculated band structure for the multilayer presented below. However, feature 1, associated with the 2DEG, does exhibit much sharper features indicative of greater spatial delocalization in the BE interval 0.0-0.4 eV.

### III.C. Standing-wave core-level photoemission and ARPES:

We next make use of the superb depth resolution of SW photoemission, including ARPES[32,33], to provide unique information on the depth distribution of the states involved in features 1 and 1'. In particular, we have used a procedure involving resonant effects, to change the scanning depth of the SW in the sample, in particular by going below and above the Gd $M_5$ absorption edge, as first pointed out in single-crystal Bragg reflection studies with x-ray fluorescence detection[34]. To validate and illustrate this new aspect of SW photoemission, we performed extensive numerical simulations with the YXRO program[35] before and after the experiment in order to choose the best photon energies and then to identify precisely the depth of origin of photoelectrons for each atomic species through core-level RCs. The RCs of different VB features were then compared to the core level RCs to provide definitive depth information. The optimization of the sample structural model was done in an iterative manner by comparing experimental and calculated rocking curves for various choices of sample configuration. The model with the best fit between experimental data and simulations is presented below.

The resonant optical constants needed for our simulations were obtained by measuring Gd $M_{4,5}$ x-ray absorption spectra (Fig. 5(a)) from total electron yield and applying a Kramers-Kronig transformation[52], with the end results for the real and imaginary parts of the index of refraction $n = 1 - \delta - i\beta$ being shown in Fig. 5(b).  In order to make the simulation as close to as possible to reality, the model included a surface contaminant layer consisting of 12 Å of adsorbed CO, to allow for the C- and O- containing species that are expected to be present, with this thickness being estimated from the combination of the C 1s relative intensity in the photoemission spectra and YXRO simulations. All interfaces were assumed to be ideally flat, with TEM images from prior work making this a very reasonable starting assumption[15].

Figs. 5(c) and 6(a) contain color-plots of electric field strength $|E^2|$ extending from vacuum into the first ca. 50 Å of our sample as a function of depth and incidence angle, for angles spanning the first-order Bragg angle at about 8.6°, and with photon energies of 1181 eV and 1187 eV just below and just above the Gd $M_5$ resonance, respectively.  These energies were chosen before the experiment from an extensive set of calculations at various photon energies to optimally place the peak of the SW as it is scanned through the Bragg condition from being above the first STO/GTO interface (Fig. 5(c)) to a position as close as possible to the interface (Fig. 6(a)).  In fact, changing the incident energy from 1181 to 1187 is found to change the sign of $\delta$ in the index of refraction (cf. Fig. 5(b)), and largely through this effect to move the SW peak downward by about 12 Å, and so by about 1/3 of the period $d_{ML} = 35.5$ Å.  Visual inspection of these two figures makes it clear that at 1187 eV, the interface is much more strongly excited.  In both of these figures, three vertical reference lines $P_1$, $P_2$, and $P_3$ at special points in the scan of the SW are indicated.  In Fig. 5, $P_1$ is where the SW maximally excites the first STO layer, $P_2$ is where the SW is optimally centered on the surface C+O layer, and $P_3$ is where the SW is a maximum somewhat above the surface. In Fig. 6, $P_1$ is where the STO/GTO interface will be most excited, $P_2$ is where the middle of the STO layer will be excited, and $P_3$. is where the surface C+O layer and the surface of the first STO layer will be preferentially excited. These angles also will be seen to represent different peak positions in the RCs to be discussed below.

Focusing first on the results for 1181 eV, we in Figs. 5(d) and 5(e) show calculated and experimental core-level RCs for all of the elements in the sample, respectively, grouped when possible such that those from a given layer are together. Note first that all of the experimental and theoretical RCs show both a Bragg peak and several Kiessig fringes, with theory well reproducing the relative intensities of these and the period of the Kiessig fringes. The fit of rocking curve amplitudes between the experiment and theory was improved somewhat by using non-resonant optical constants[53] for the unit cells of GTO directly adjacent to STO, similarly to the case of LaSrMnO₃/SrTiO₃[32]. The change of the optical properties in the interfacial GTO layer can be due to a different Gd $3d_{5/2}$ resonant behavior, possibly connected to the 2DEG formation in STO and a charge transfer between GTO and STO. The rocking curve shown for O 1s originates in both layers in the multilayer. However, due to the inelastic attenuation of photoelectron intensities, the signal from the top-layer STO is expected to dominate the O1s spectra, and in fact, the shape and phase of the O 1s RC coincides with the one for Sr 3d. C 1s comes from the surface contaminant layer, and its RC is shifted considerably from all other RCs, due to its unique depth distribution. As might be expected from the different alternating depth distributions of Gd and Sr atoms in the sample, the rocking curves of Sr 3d and Gd 4f show opposite phases as the SW scans through them, demonstrating the high sensitivity of such SW RC measurements to the depth of a given species. Beyond this, we can make use of the two components in the Ti 2p spectrum shown in Fig. 2(e) to try to depth-resolve the positions of these two types of Ti. RCs were thus separately determined for Ti $2p^{4+}$ and Ti $2p^{3+}$ by peak fitting, and these are shown separately in Fig. 5(e). We find that the Sr 3d and Ti $2p^{4+}$ RCs are almost identical in both experiment and theory, including the Kiessig fringes that are evident for angles below the Bragg angle, as they should be in originating uniquely in the STO layers of the sample. The same is true of Ti $2p^{3+}$ and Gd 4f, which are expected to originate in the GTO layers of the sample. Although the RC for Ti $2p^{4+}$ has a slightly different shape from Gd 4f, this is largely due to the greater difficulty of fitting this weaker peak. The average $Ti^{3+}/Ti^{4+}$ intensity ratio of ~7 over the RCs, and measured at both 1181 and 1187 eV before and after the Gd M5 edge is also qualitatively consistent with the total relative number of these ions in our sample, which is ½, provided that we also allow for the fact that STO is the top layer and is thus enhanced due to photoelectron inelastic scattering; this conclusion has been confirmed using the SESSA program for simulating XPS spectra[30]. An experimental RC for the LHB determined by integrating all intensity over regions B, C, and D as defined in Fig. 4(b) in a spectrum taken in the room-temperature MEWDOS limit yields the curve shown at the bottom of Fig. 5(e).  Although noisier than the others due to the low relative intensities of this VB feature (cf. Figs. 2(a)-(b) and 4(a)-(b)) and the difficulty in separating it from underlying inelastic background, the form of the LHB RC is at least qualitatively consistent with its arising throughout the GTO if one notes its strong similarity to the Gd 4f RC, or compares experiment to theory in Figs. 5(c)-(d).  These RCs at 1181 eV thus already provide direct information concerning the depth distributions of different chemical species and the LHB.

Finally, the LBE and HBE components of Sr 3d in Fig. 2(f) are found via peak fitting intensity analysis to have the essentially identical RC shapes shown in Fig. 5(e), and to be very similar to the RC for $Ti^{4+}$, suggesting possibly two different bonding or charge states[42], or a screening satellite mechanism that mixes HBE = $Sr^{2+}$ and LBE screened $Sr^{2+} \approx Sr^{+1}$ in the final state, <u>but with either mechanism being distributed through the entire STO layer</u>.  These results are not consistent with a valence-band offset at each interface that decays via band bending through the STO layer, as this situation would result in different depth distributions and RC forms for the two peaks.  Additional data we have obtained for a trilayer system of STO sandwiched between GTO (not shown here)[54], in which no bare-surface effects are possible for STO, further reveal a direct correlation between the presence of the 2DEG and the relative strength of the LBE component, with these results favoring the screening mechanism.  A simple Anderson Impurity Model calculation based only on localized screening with charge transfer into Sr 4d yields a splitting between the two components that is too large, at about 6.4 eV[40], but a prior theoretical study of screening into Sr 5s based on a point-ion plus Madelung potential model yields a value of 1.6 eV that is much closer to experiment[39].  An additional factor here could be the involvement of the highly delocalized electronic states responsible for the 2DEG in the screening, as has been seen e.g. in transition-metal 2p spectra at higher energies[55], postulated recently for transition-metal 2p spectra in other oxides[56], and also discussed for La 3d spectra in several compounds[57].  It is also possible that these two states could be due to exciton-like effects[58].  More theoretical calculations will be needed to confirm the final-state screening model for Sr 3d, but the SW results shown here definitely point to an intriguing new aspect of the Sr 3d spectroscopy that is in turn directly related to the spatial distribution of the 2DEG.

Figs. 6(c) and 6(d) now show the same kind of theoretical and experimental RC comparisons for an excitation energy of 1187 eV that focuses the SW more on the interface region and again for angle- integrated low temperature VB results that approximate the MEWDOS limit.  From the top down, the O 1s, C 1s, Sr 3d, and Gd 4f RCs are quite different from those at 1181 eV, and again in excellent agreement with theory, including the relative intensity and spacing of the Kiessig fringes.  This degree of agreement further confirms the accuracy of the resonant optical constants and the validity of the structure we are assuming for the sample.  It also further demonstrates our previously established SW calculation methodology[32,33,35]. The RCs of Sr 3d and Gd 4f (as well as Gd 4d, not shown here) again strongly differ, with maxima and minima out of phase, a direct consequence of their origin from different layers of the sample. Our simulation predicts this behavior correctly, including the amplitude of the standing wave effect, which, as measured by maximum-minus-minimum, is almost 30% for both of these core-levels. Comparing further the experimental RC for Peak 1 near $E_F$ (which could be resolved in these results more quantitatively due to better statistics than those at 1181 eV) and the theoretical curve for a 2DEG assumed to be spread over the entire thickness of the STO layer, we see excellent agreement in shape, indicating that peak 1 originates in STO, but by the strong dissimilarity with the Gd 4f (or Gd 4d) RCs, that peak 1 is not associated with GTO. An additional theoretical simulation with the 2DEG Peak 1 localized in STO but within 4 Å (~1 u.c. of STO) of the STO/GTO interface (dashed violet curve) strongly disagrees with experiment, thus further confirming our conclusion concerning the spatial extent of the 2DEG. In Fig. 6(b), we further show a systematic variation of the thickness assumed for the 2DEG, and this illustrates the high sensitivity of the data to the 2DEG spatial distribution. The fact that the C 1s and Gd 4f rocking curves do not vary in theory is simply that dividing the STO up into interface and non-interface regions has no effect on their interaction with the SW. The total dissimilarity of the C 1s experimental and theoretical RCs to those for peak 1 and the interface in Fig. 6(c) also rule out the presence of a surface 2DEG on STO, even though such a state has in fact been observed on specially prepared clean STO surfaces[59,60]. The LHB RC, while again noisier for reasons noted above, nonetheless agrees semi-quantitatively with those of Gd 4f, including the low-angle Kiessig fringes, and with a curve calculated assuming the LHB extends throughout the GTO. These standing-wave RC results thus provide a clear confirmation that Peak 1 represents the 2DEG and that, as far as the intensity in that energy regime is concerned, the origin of this peak extends throughout the entire STO layer. Again, the LBE and HBE components found for Sr 3d in Fig. 2(f) are found to have essentially identical shapes, suggesting at least two different bonding or charge states, or a screening satellite mechanism that must be identical throughout the entire STO layer, as discussed above.

We now consider in Fig. 7 more detailed experimental ARPES results obtained at 1181 eV excitation and for an incidence angle away from any SW formation, with panel 7(a) showing the sampling in **k**-space as determined by direct transitions (DTs) satisfying $\mathbf{k}_i = \mathbf{k}_f - \mathbf{k}_{h\nu} - \mathbf{g}$, where $\mathbf{k}_i$ is the initial state wave vector, $\mathbf{k}_f$ is the final-state photoelectron wave vector inside the surface, assumed in this drawing to be a free-electron state, $\mathbf{k}_{h\nu}$ is the photon wavevector and $\mathbf{g}$ is the relevant reciprocal lattice vector supporting the transition; 7(b) some raw ARPES data in $k_x$-$k_y$ over the near-$E_F$ region of feature 1, superimposed upon a projected cross section of the sampling sphere, and 7(c)-(e) several representations, including cross section cuts, of the three-dimensional dataset in $k_x$, $k_y$, and *BE*.  The small correction due to photon momentum vector is indicated in both 7(a) and 7(b).  The horizontal top surface in 7(c) resembles that in 4(c)(iii)-(iv) for emission from the A and B energy regions, with maxima at the $\Gamma$ point and intensity sleeves connecting those maxima through *X* points for this energy.  The perpendicular cuts in 7(d) and 7(e) reveal new details of the electronic structure, with distinct maxima near the $\Gamma$ points for the energy range of peak 1 and much more diffuse structure in ***k*** for the energy range of peak 1', although 7(e) in fact shows to a higher degree the similar momentum intensity distribution of peaks 1 and 1' over a broader *BE* range down to ~4 eV.  Thus, although these two features show very much the same pattern in $k_x$-$k_y$ ARPES cross sections, and thus have led us to suggest them to be strongly admixed, they nonetheless have very different intensity profiles when viewed in these $k_x$-*BE* or $k_y$-*BE* cross sections.  Beyond this, 7(d) and 7(e) reveal additional dispersive features in the electronic structure over *BEs* of ~4-9 eV that we compare to theory below.  The non-dispersive nature of the bands over *BEs* of ~10-12 eV identifies them as being highly localized in character, and thus exhibiting only XPD effects, supporting our previous suggestion that they must represent Gd 4f bands with no dispersion.

Figs. 8(a)-(c) at a non-resonant 833 eV excitation are thus lower-energy analogues of Figs. 7(c)-(e) at resonant 1181 eV excitation, and they fully support the conclusions based on Fig. 7.  The top surface in 8(a), analogous to the one in 7(a), is in fact identical to the plot in Fig. 4(c)(i).  As noted in connection with Fig. 4(c)(i), the energy used here and the larger acceptance angle of the spectrometer permits observing a much larger number of BZs.

Based on our combined experimental results including MEWDOS VB results, resonant ARPES, and standing-wave core- and VB- photoemission, we thus conclude that Peak 1 represents the 2DEG formed by the presence of the STO/GTO interfaces, that it is spatially localized within the STO layers, that it exhibits non-uniform momentum distribution very similar to Peak 1' which we assign to the LHB of GTO, but that it is much more localized in ***k*** than the LHB.  It is also worth noting that our results for this buried interface do not appear consistent with the suggestion in prior 2DEG photoemission studies involving either a free surface of STO[59,60], or a buried interface of STO[6], of the essential role played by oxygen vacancies in forming both the 2DEG and states at the position of peak 1', in that is seems very unlikely that these would be uniformly distributed over both the STO and GTO layers, as our standing-wave analysis indicates.

**III.D. More detailed comparisons to theory:**

In order to better understand our results, we now consider theoretical calculations at several levels.  Figure 9(a) depicts the calculated band structure of the multilayer, clearly indicating its metallic character with electrons residing in STO subbands. The figure focuses on the energy region near $E_F$ encompassing the LHB and the 2DEG for bands along *X-Γ-M/2,* again exhibiting ferromagnetic order such that the spin-up and spin-down bands are non-degenerate. The fact that the electrons reside in the STO is further confirmed by Figs. 9(b) and (c). Figure 9(b) shows via charge-density plots that the electronic charge density associated with the STO subbands is entirely located within the STO, and fairly uniformly distributed over all the atomic layers within the STO. In Fig. 9(c) we further show the calculated atomic layer-by-atomic layer depth-resolved charge densities of Ti-derived states in the LHB (as integrated over BEs of 0.53 to 1.65 eV and shown as the orange curves), and in the 2DEG (as integrated over 0.00 to 0.52 eV and shown as the red curves). As expected, the LHB resides entirely in the GTO, and this is consistent with our SW results. Again we observe that the charge density of the states making up the 2DEG is relatively constant through the entire STO layer, peaking of course at each $TiO_2$ layer, and showing a slight splitting at the middle of the layer due to the presence of greater $t_{2g}$-$d_{yz}$ and $t_{2g}$-$d_{xz}$ character there, compared to the case near the interface, where there is greater $t_{2g}$-$d_{xy}$ character. An integration over the charge density residing in the STO subbands produces a value of 1 electron, consistent with having two identical STO/GTO interfaces, with each contributing 1/2 electrons per 5-atom unit-cell area. These theoretical results provide further evidence that oxygen vacancies are not required to produce the type of multilayer 2DEG that our standing-wave results have clearly determined. The 2DEG is truly an intrinsic feature of the interface.

In Fig. 9(d) we make further contact with the ARPES results by comparing the complex sets of bands resulting from the hybrid-functional DFT calculation to the experimental $k_y$-BE ARPES plot from Fig. 7(d). Even though this kind of comparison neglects any sort of k-conservation or more complex matrix-element effects, there is in general excellent agreement as to where one expects to see intensity, including the more localized nature of the 2DEG in ***k***. The density of bands connected to the 2DEG around the central *Γ* point in Fig. 9(d) compared to the two *Γ* points in the adjacent Brillouin zones is smaller due to their different **$k_z$** component and the different proximity to the true center of the BZ for each of them (cf. Fig. 7(a)).

We finally return to the MEWDOS level of looking at our results and compare theory to angle-integrated experimental results with 833 eV excitation. This is done in Fig. 10, where the orbital-projected densities of states in the multilayer, including of course the LHB and the 2DEG, have been multiplied by the respective differential atomic photoelectric cross section for each orbital, and the sum of all contributions is compared to our experimental results. The theory has been truncated at $E_F$, convoluted with the Fermi function for 300 K over that energy region, and then convoluted with a Gaussian of 0.30 eV to simulate the overall experimental resolution. The agreement as to the main features in experiment and theory is very good, including for the 2DEG and the LHB shown in the blowup of Fig. 10(b). Of course, the Gd 4f bands and their features in the experiment are missing from theory, where these states are treated as core levels, as noted previously. As a further comment here, the observation that the experimental relative intensities of the LHB and the 2DEG are qualitatively similar to those in theory in Fig. 10(b), although with the LHB weaker in experiment as judged by relative areas, is due to the fact that the experiment samples states through both the top STO layer and also from the buried GTO layer. More quantitatively, the estimated IMFPs of 22(20) Å in STO(GTO) mentioned above indicate that the LHB should be attenuated more by the overlying STO layer. A more accurate way to compare these relative intensities would be to allow for the precise depth distribution of both the LHB states in the GTO layers and the 2DEG states in the STO layers, and incorporate inelastic scattering, but even at the level of Fig. 10, theory and experiment are in very good agreement.

As a final aspect of theory, we have carried out one-step photoemission calculations in the relativistic spin-polarized KKR formalism for the multilayer[38], so as to include matrix element effects beyond $k$-conservation. These calculations allowed for the tetrahedral distortions present in GTO that result in the larger unit cell of Fig. 3(b). Some of these results for states at the Fermi energy are shown in Fig. 4(c)(viii) as Theory. These calculated patterns are very similar to experiment in showing maxima near $\Gamma$ and intensity stretching along $\Gamma$-$X$ directions between the maxima. Even though they do not fully reproduce the simple square-array pattern of experiment, they are generally consistent with it. We also note some weak intensity in the centers of the square-array features that may represent a small effect of the larger unit cell of Fig. 3(b), but seen only weakly in the ARPES, something we have discussed above. These one-step calculations thus further assist in confirming the character of the 2DEG electronic structure.

### IV. CONCLUSIONS AND FUTURE OUTLOOK

By using an array of photoemission techniques including soft and hard x-ray excitation, core-level spectroscopy, soft x-ray ARPES, resonant excitation, and standing-wave effects enhanced by resonant tuning below and above absorption edges, combined with state-of-the-art theoretical calculations of the electronic structure and the photoemission process, we have thus characterized the spatial and momentum properties of the 2DEG formed at the STO/GTO interface, determining via standing-wave depth-distribution measurements that it resides throughout the 6 unit-cell STO layer, a result consistent with prior experiment and theory for 2DEGs in STO, but via a more direct measurement[14], and measuring the momentum dispersion of its states. Core-level binding shifts and other fine structure, e.g. of Ti 2p and Sr 3d, assist with this analysis, and suggest a novel interpretation of the spectra observed, particularly for final-state screening in Sr 3d. Our experimental results are furthermore supported in several ways by theory, including both ground state and one-step photoemission calculations, and lead to a much more complete picture of the origin and nature of this 2DEG. Charge transfer at the GTO/STO interfaces is also found to be the cause of the 2DEG, rather than oxygen vacancies as suggested in connection with other 2DEG observations[6,27]. We suggest that similar multi-technique photoemission studies of such states at buried interfaces, combined with theory at the level we have employed, will be a very fruitful future approach for exploring and modifying the fascinating world of 2DEG and buried-interface physics, with applications in a wide variety of fields in materials science, physics, chemistry, and applied science. In fact, our above- and below- resonance SW technique has in more recent work been successfully applied to the system of $LaCoO_3$ and $SrTiO_3$[61], two non-ferroelectric materials which have been found to exhibit ferroelectric behavior at their interface[62]. As other examples looking ahead, we note the recent application of soft x-ray standing-wave photoemission to the liquid-solid interface[13,63], with the above- and below- resonance methods described here promising to yield much more precise characterization of such interfaces, in electrochemical cells[63] or gated devices involving a solid oxide and a liquid electrolyte[10].

### ACKNOWLEDGEMENTS

Primary support for this work is from the MURI program of the Army Research Office (Grant No. W911-NF-09-1-0398). The Advanced Light Source, A.B., W.C.S., and C.S.F. are supported by the Director, Office of Science, Office of Basic Energy Sciences, Materials Sciences and Engineering Division, of the U.S. Department of Energy under Contract No. DEAC02-05CH11231. P.M. was supported by the U.S. National Science Foundation (Grant No. DMR-1006640). The HXPS measurements at BL15XU of SPring-8 were performed under the approval of NIMS Beamline Station (Proposal No. 2011A4606). The HAXPES instrument at Petra III beamline P09 is jointly operated by the University of Wurzburg (R. Claessen), the University of Mainz (C. Felser), and DESY. Funding by the Federal Ministry of Education and Research (BMBF) under contracts 05KS7UM1, 05K10UMA, 05KS7WW3, and 05K10WW1 is gratefully acknowledged. A.J. and C.G.V.d.W. were supported by the US Army Research Office (W911-NF-11-1-0232) and L.B. by the NSF MRSEC Program (DMR-1121053). Computational resources were provided by the XSEDE supported by NSF (ACI-1053575 and DMR07-0072N). S.N. received support in the completion of this work from the Jülich Research Center. GKP also thanks the Swedish Research Council for financial support. A.R. was funded by the Royal Thai Government and C.C. was funded by GAANN program through UC Davis Physics Department. C.S.F. has also been supported during the writing of this paper for salary by the Director, Office of Science, Office of Basic Energy Sciences, Materials Sciences and Engineering Division, of the U.S. Department of Energy under Contract No. DE-AC02-05CH11231, by the Laboratory Directed Research and Development Program of Lawrence Berkeley National Laboratory under the same contract, by Contract No. DE-SC0014697 from DOE Office of Basic Energy Sciences, Materials Science and Engineering Division through the University of California Davis, and by the LabEx PALM program "Investissements d'Avenir"overseen by the French National Research Agency (ANR) (reference: ANR-10-LABX-0039). One-step theory calculations of J.M., J.B. and H.E. are supported by Bundesministerium für Bildung und Forschung Project 05K13WMA, and also acknowledges support by the CENTEM (CZ.1.05/2.1.00/03.0088) and CENTEM PLUS (LO1402), co-funded by the European Regional Development Fund, through the Ministry of Education, Youth and Sports of the Czech Republic OP RDI Programme.

**FIGURE CAPTIONS:**

**Figure 1.** (a) The experimental geometry and the structure of the sample, with various key parameters defined. (b) A layer-by-layer drawing of a single GTO/STO bilayer in the sample, with 3 unit cells (u.c.) of GTO and 6 u.c. of STO.

**Figure 2.** Valence-band (VB) and core-level spectra with soft- and hard- x-ray excitation and at room temperature representing matrix-element weighted densities-of-state (MEWDOS) for the valence spectra. (a) The full VB spectrum with 833 eV excitation. Various peaks and shoulders in the spectrum are labelled, from 1, 1'…5. (b) A blowup of the near-EF region of (a). (c) Full VB spectra with hard x-ray excitation for bulk GTO, bulk STO, and the multilayer sample. (d) A blowup of the near-$E_F$ region in (c). (e) The Ti 2p spectrum from the multilayer with 833 eV excitation, exhibiting two components that are attributed to Ti present in 4+ and 3+ states. (f) The Sr 3d spectrum from the multilayer, with two possible charge or potential states being suggested by the curve fitting indicated. Data obtained at (a), (e), and (f)-ALS, and (c), (d)- ALS, SPring-8, and Petra III.

**Figure 3.** First-principles results for electronic structure of bulk GTO and bulk STO. (a) Band structure along high-symmetry lines in the Brillouin zone for bulk GTO with the unit cell shown in (b). The locations of the lower and upper Hubbard bands (LHB and UHB) are indicated. (c) Densities of states for bulk GTO: Total, with 0.30 eV convolution to mimic experimental resolution, and orbital-projected for Ti, O, and Gd. (d) Superposition of the DOS for GTO and STO, aligned according to the calculated band offset of 2.6 eV.

**Figure 4.** Resonant photoemission from the [6 u.c. STO/3 u.c. GTO]x20 multilayer sample. (a) The effect of scanning photon energy through the Ti 2p absorption resonances on MEWDOS-level VB spectra. Labelled features are: pre-edge, a = Ti $2p_{3/2}$-3d $t_{2g}$, b = Ti $2p_{3/2}$-3d $e_g$, c = Ti $2p_{1/2}$-3d $t_{2g}$, and d = Ti $2p_{1/2}$-3d $e_g$. (b) Resonant MEWDOS VB spectrum near $E_F$ with 465.2 eV excitation. The features 1 and 1' are indicated, together with five energy regions A-E. (c) Non-resonant and resonant kx-ky

ARPES maps for: (i) Non-resonant excitation at 833 eV, region A, (ii) Non-resonant excitation at 470 eV, region A, (iii) Resonant excitation at 465.2 eV, region A, (iv)-(vii) Resonant excitation at 465.2 eV for regions B-E, respectively, and (viii) theoretical prediction for region A from one-step photoemission theory. All data collected at ALS.

**Figure 5.** Standing-wave photoemission from the [6 u.c. STO/3 u.c. GTO]$_{x20}$ multilayer sample with excitation at 1181 eV. (a) X-ray absorption spectrum measured through the Gd $M_5$ resonance. (b) The variation of the real and imaginary parts of the index of refraction ($\delta$ and $\beta$) with the two energies selected for standing-wave photoemission indicated. (c) The electric field strength $|E^2|$ as a function of depth and incidence angle. (d) Calculated RCs for various core levels and the LHB, assuming the structure in (c). (e) Experimental RCs for the same core levels, but with LBE and HBE Sr 3d shown separately, and the LHB. Experimental data are from the ALS.

**Figure 6.** Figure 6. Standing-wave photoemission from the [6 u.c. STO/3 u.c. GTO]$_{x20}$ multilayer sample with excitation at 1187 eV. (a) The calculated electric field strength $|E^2|$ as a function of depth and incidence angle. (b) X-ray optical calculations (solid lines) of three of the key rocking curves in Figure 6(d) (empty circles), with variation only of the thickness into which the STO layers are divided up, showing through the 2DEG that filling the full thickness of STO with the 2DEG is the optimum result. The C 1s and Gd 4f RCs are not influenced by the division of STO into interface and non-interface regions. (c) Calculated RCs for various core levels, the 2DEG, and the LHB, assuming the structure in (a), and for theory assuming the 2DEG to be spread throughout the STO (solid turqoise curve) and localized within 4 Å of the GTO (dashed violet curve). (d) Experimental RCs for the same core levels, but with the Sr 3d HBE and LBE shown separately, and the LHB. Experimental data are from the SLS.

**Figure 7.** Non-resonant ARPES results at 1181 eV excitation from the multilayer sample. (a) The region in $k_x$-$k_z$ sampled, with the red arc as seen by the detector, and the $k$-conservation in a direct transition being indicated. (b) A $k_x$-$k_y$ ARPES map over energy interval A in Fig. 4(b) encompassing feature 1 near $E_F$. (c) A partial representation of the 3D ARPES dataset in $k_x$, $k_y$ and BE. (d) A cross section of the data of (b) in $k_y$-BE. (e) A cross section of the data in (b) in $k_x$-BE. Experimental data are from the SLS.

**Figure 8.** As Figures 7(c)-(e), but for a non-resonant excitation energy of 833 eV, and with a larger spectrometer acceptance angle that leads to sampling a greater number of BZs. Experimental data are from the ALS.

**Figure 9.** Theoretical calculations for the electronic structure of the multilayer sample, including a comparison to experiment. (a) Energy bands along the X-Γ-M/2 directions, with the STO subbands, including the 2DEG and the LHB, indicated. (b) Charge densities of the occupied electronic states, as isosurfaces plotted at 10% of maximum value: STO conduction bands in yellow, oxygen atoms in red, strontium in white, titanium in blue and gadolinium in black. (c) Depth-dependent in-plane averaged charge densities. The central TiO2 layer is underlined in yellow, as in Fig. 1(b). The orange curves correspond to the LHB energy range (in GTO), and the red curves to the 2DEG (in STO). The 2DEG charge density integrates out to the expected ½ electron per unit cell area per each of the two interfaces, or an electron density of $3 \times 10^{14}$ cm$^{-2}$. (d) Comparison of a nonresonant $k_x$-BE ARPES map at 1181 eV with calculated bands along Γ-X-Γ-X-Γ, and over a more extended BE range than in (a).

**Figure 10.** (a) The experimental MEWDOS VB spectra from the multilayer at 833 eV (Fig. 2(a),(b)) are compared to a simulated spectrum constructed by summing orbital-projected densities of states for the multilayer that are each multiplied by the appropriate differential photoelectric cross section. The Gd 4f states were not included in the calculation. (b) Enlargement of the region near $E_F$.

## Figure 1 (color; 1-column)

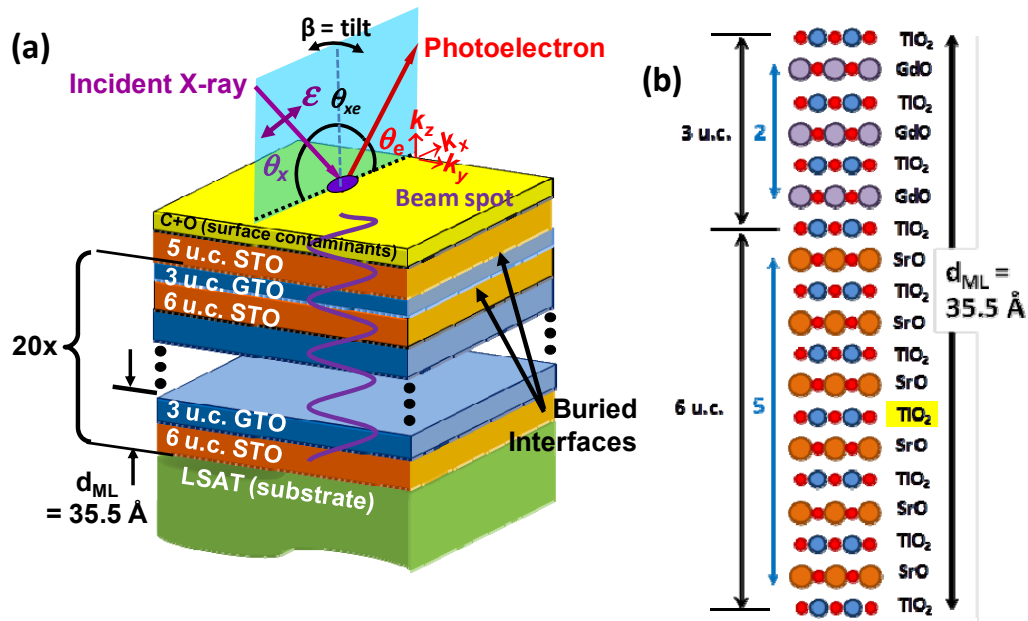

Figure 1. (a) The experimental geometry and the structure of the sample, with various key parameters defined. (b) A layer-by-layer drawing of a single GTO/STO bilayer in the sample, with 3 unit cells (u.c.) of GTO and 6 u.c. of STO.

**Figure 2 (color; 1-column)**

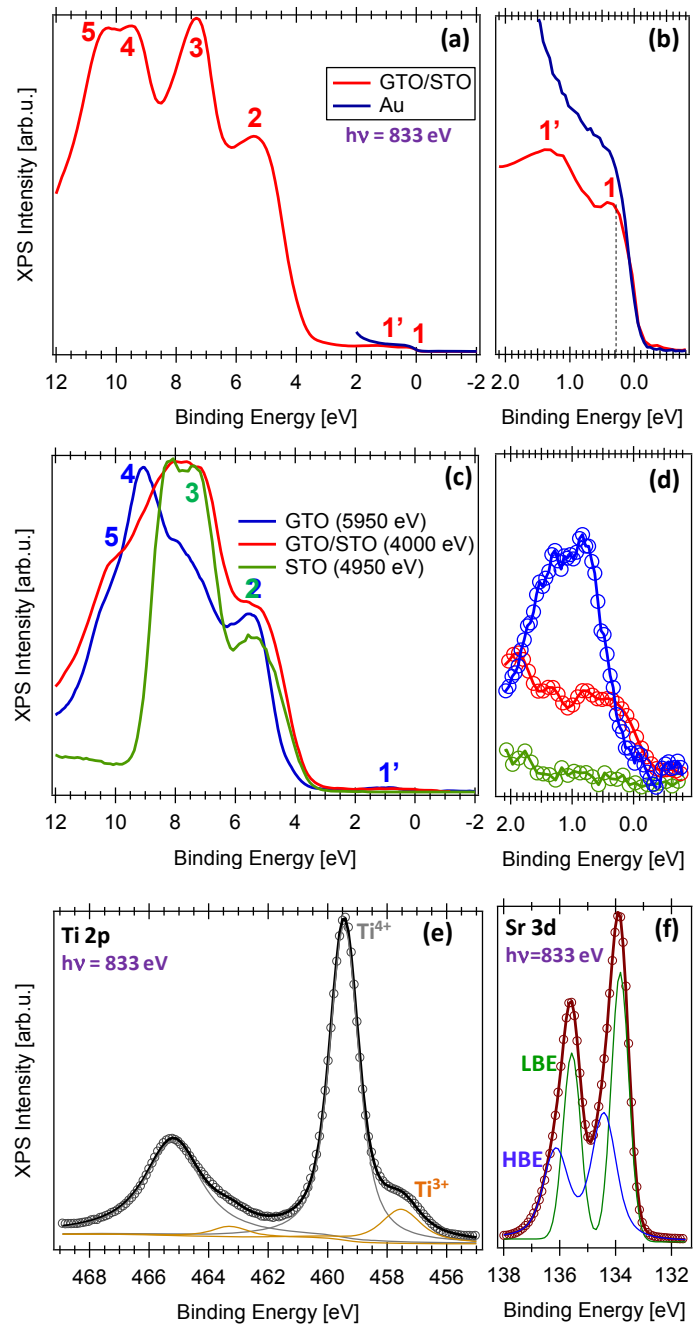

Figure 2. Valence-band (VB) and core-level spectra with soft- and hard- x-ray excitation and at room temperature representing matrix-element weighted densities-of-state (MEWDOS) for the valence spectra. (a) The full VB spectrum with 833 eV excitation. Various peaks and shoulders in the spectrum are labelled, from 1, 1'...5. (b) A blowup of the near-$E_F$ region of (a). (c) Full VB spectra with hard x-ray excitation for bulk GTO, bulk STO, and the multilayer sample. (d) A blowup of the near-$E_F$ region in (c). (e) The Ti 2p spectrum from the multilayer with 833 eV excitation, exhibiting two components that are attributed to Ti present in 4+ and 3+ states. (f) The Sr 3d spectrum from the multilayer, with two possible charge or potential states being suggested by the curve fitting indicated. Data obtained at (a), (e), and (f)-ALS, and (c), (d)- ALS, SPring-8, and Petra III.

**Figure 3 (color; 1.5-column)**

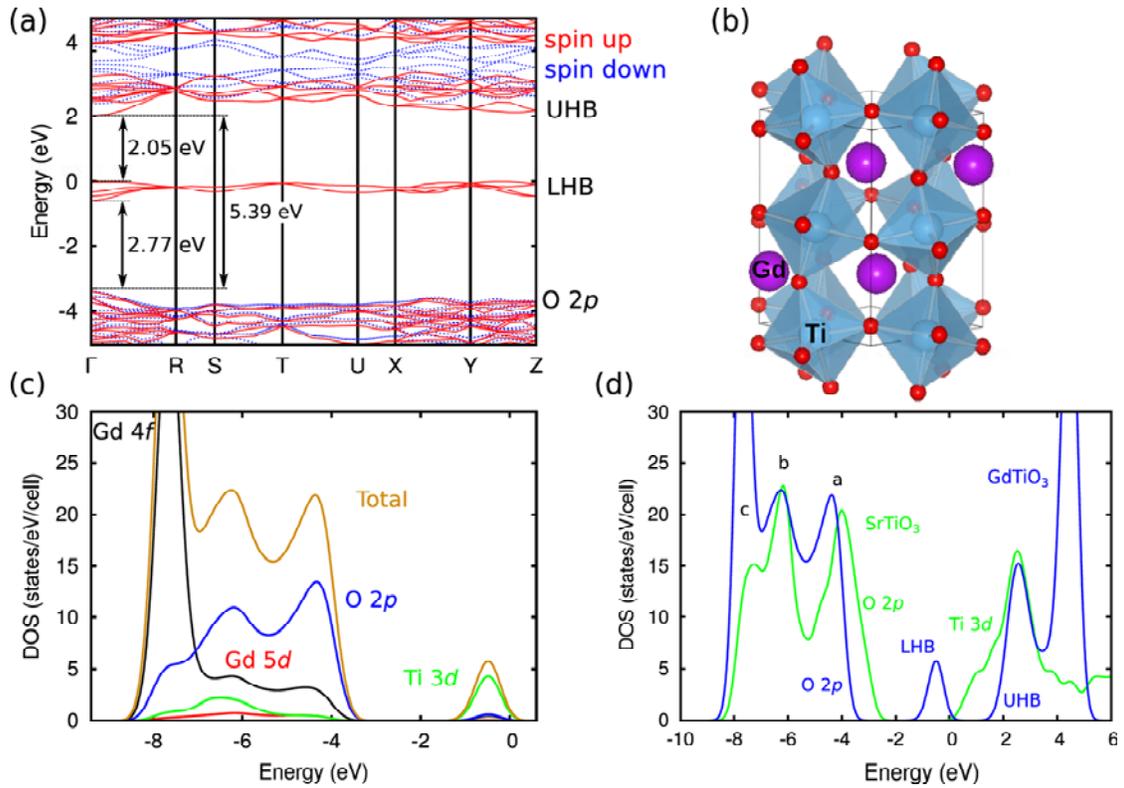

**Figure 3. First-principles results for electronic structure of bulk GTO and bulk STO. (a) Band structure along high-symmetry lines in the Brillouin zone for bulk GTO with the unit cell shown in (b). The locations of the lower and upper Hubbard bands (LHB and UHB) are indicated. (c) Densities of states for bulk GTO: Total, with 0.30 eV convolution to mimic experimental resolution, and orbital-projected for Ti, O, and Gd. (d) Superposition of the DOS for GTO and STO, aligned according to the calculated band offset of 2.6 eV. Zero energy is set to the top of the LHB.**

**Figure 4 (color; 1.5column)**

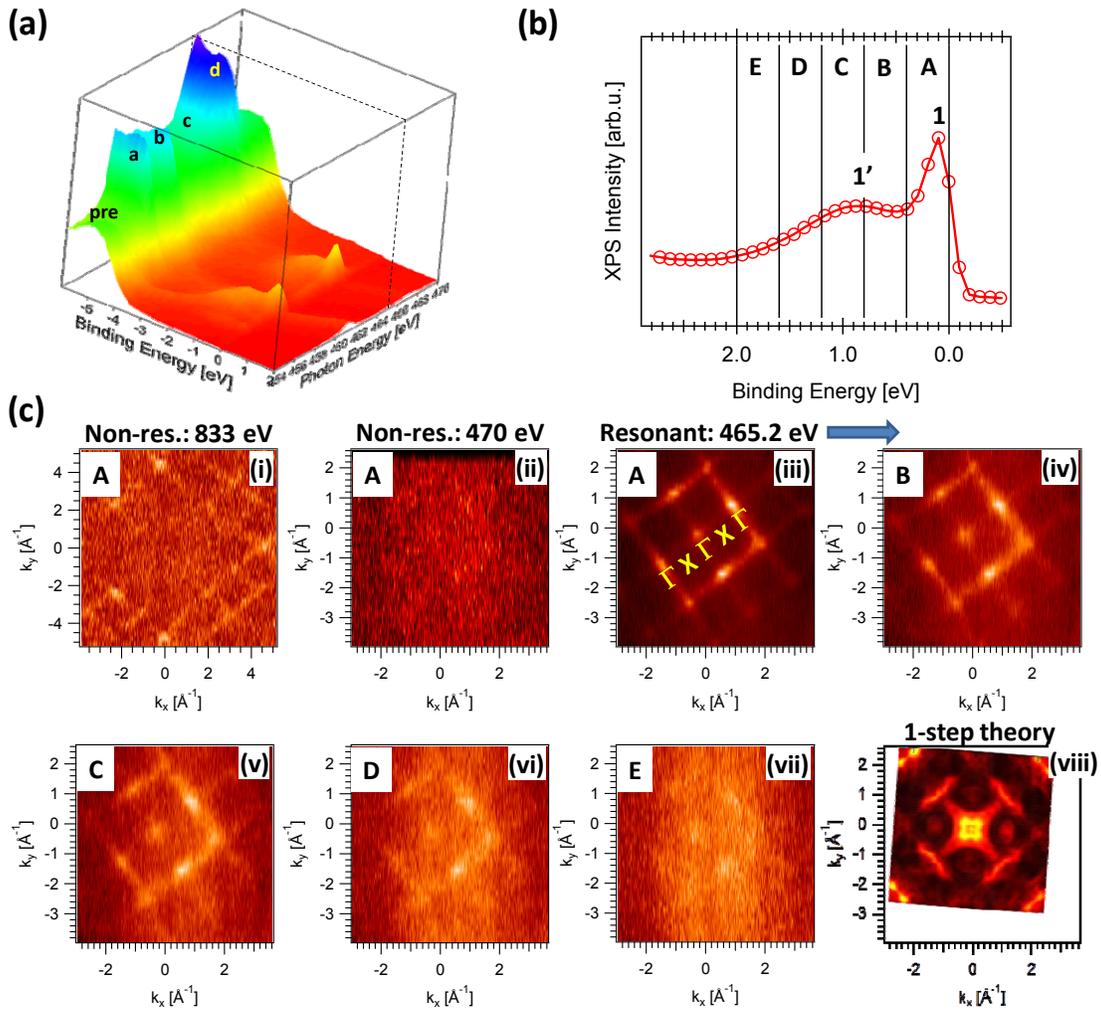

Figure 4. Resonant photoemission from the [6 u.c. STO/3 u.c. GTO]x20 multilayer sample. (a) The effect of scanning photon energy through the Ti 2p absorption resonances on MEWDOS-level VB spectra. Labelled features are: pre-edge, a = Ti $2p_{3/2}$-3d $t_{2g}$, b = Ti $2p_{3/2}$-3d $e_g$, c = Ti $2p_{1/2}$-3d $t_{2g}$, and d = Ti $2p_{1/2}$-3d $e_g$. (b) Resonant MEWDOS VB spectrum near $E_F$ with 465.2 eV excitation. The features 1 and 1' are indicated, together with five energy regions A-E. (c) Non-resonant and resonant $k_x$-$k_y$ ARPES maps for: (i) Non-resonant excitation at 833 eV, region A, (ii) Non-resonant excitation at 470 eV, region A, (iii) Resonant excitation at 465.2 eV, region A, (iv)-(vii) Resonant excitation at 465.2 eV for regions B-E, respectively, and (viii) theoretical prediction for region A from one-step photoemission theory. The theoretical ARPES map has been rotated slightly to agree with a slight azimuthal rotation in experiment. All data collected at ALS.

Figure 5 (color; 1.5-column)

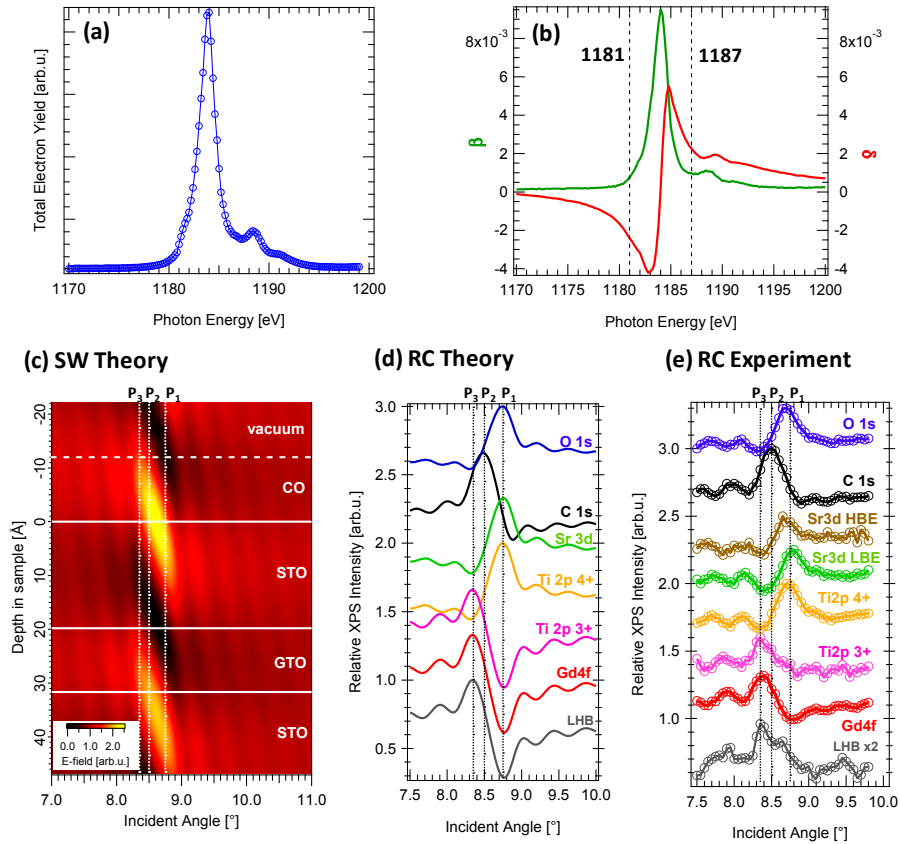

**Figure 5.** Standing-wave photoemission from the [6 u.c. STO/3 u.c. GTO]x20 multilayer sample with excitation at 1181 eV. **(a)** X-ray absorption spectrum measured through the Gd M5 resonance. **(b)** The variation of the real and imaginary parts of the index of refraction ($\delta$ and $\beta$) with the two energies selected for standing-wave photoemission indicated. **(c)** The electric field strength |E²| as a function of depth and incidence angle. **(d)** Calculated RCs for various core levels and the LHB, assuming the structure in (c). **(e)** Experimental RCs for the same core levels, but with LBE and HBE Sr 3d shown separately, and the LHB. Experimental data are from the ALS.

# Figure 6 (color; 1-column)

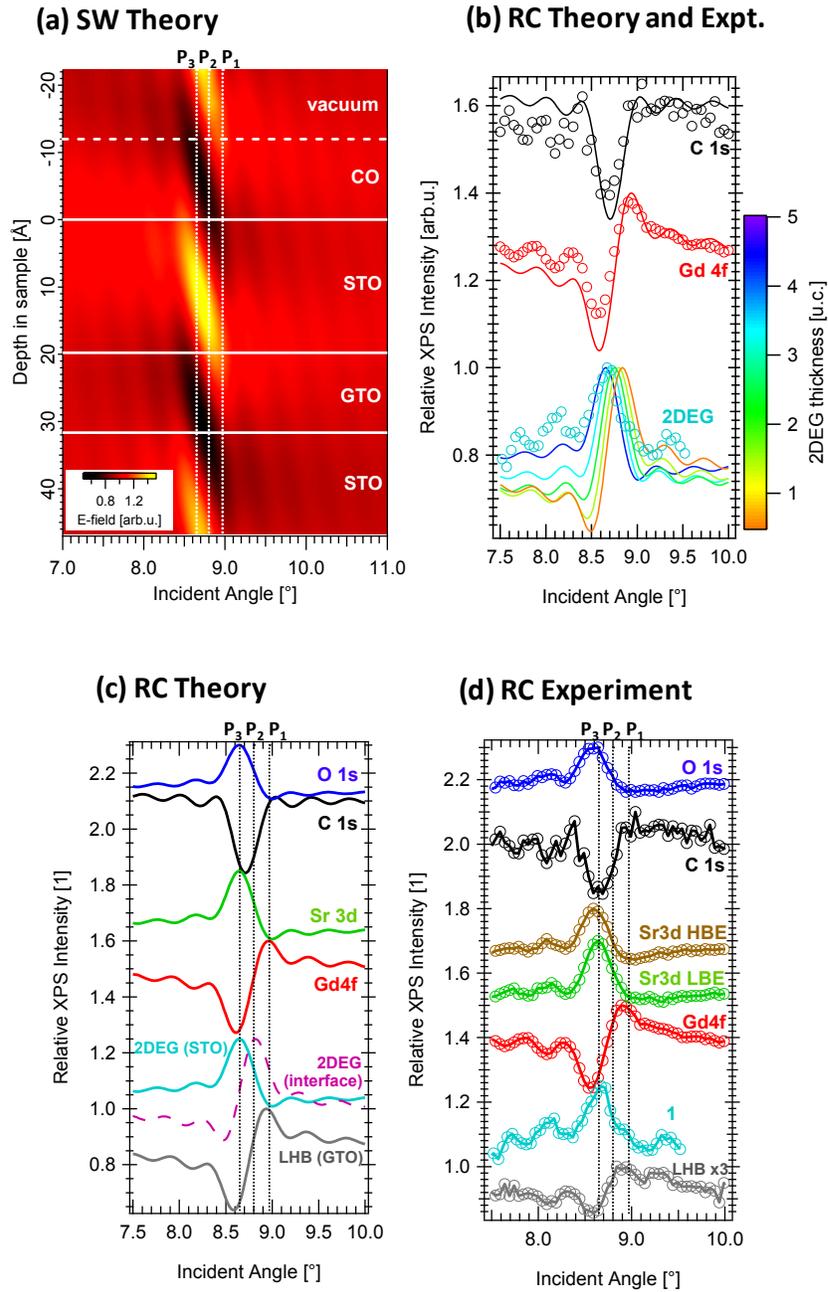

**Figure 6.** Standing-wave photoemission from the [6 u.c. STO/3 u.c. GTO]$_{×20}$ multilayer sample with excitation at 1187 eV. (a) The calculated electric field strength |E²| as a function of depth and incidence angle. (b) X-ray optical calculations (solid lines) of three of the key rocking curves in Figure 6(d) (empty circles), with variation only of the thickness into which the STO layers are divided up, showing through the 2DEG that filling the full thickness of STO with the 2DEG is the optimum result. The C 1s and Gd 4f RCs are not influenced by the division of STO into interface and non-interface regions. (c) Calculated RCs for various core levels, the 2DEG, and the LHB, assuming the structure in (a), and for theory assuming the 2DEG to be spread throughout the STO (solid turqoise curve) and localized within 4 Å of the GTO (dashed violet curve). (d) Experimental RCs for the same core levels, but with the Sr 3d HBE and LBE shown separately, and the LHB. Experimental data are from the SLS.

**Figure 7 (color; 2-column)**

**(a)** [001] [100] $\theta_x = 20°$ $2\pi/a = 1.59$ Å$^{-1}$ 1.59 Å$^{-1}$ $\vec{g}$ $\Gamma$ X X $k_{h\nu}$ X $\Gamma$ X $\vec{k_f}$ $-\vec{k_{h\nu}}$ $h\nu = 1181$ eV $\vec{k_f} - \vec{k_{h\nu}}$ $|k_f| = 17.61$ Å$^{-1}$ $|k_{h\nu}| = 0.60$ Å$^{-1}$ Γ-X-Γ = 1.59 Å$^{-1}$

**(b)** [010] [100] $-k_{h\nu}$ ● = surface normal

**(c)** $k_y$ $k_x$ Γ Γ X Γ X Γ Binding Energy [eV]

**(d)** Γ X Γ X Γ $E_F$ $I_1$ ~O2p Gd 4f Binding Energy [eV] $k_y$ [Å$^{-1}$]

**(e)** Γ X Γ $E_F$ Binding Energy [eV] $k_x$ [Å$^{-1}$]

**Figure 7. Non-resonant ARPES results at 1181 eV excitation from the multilayer sample. (a)** The region in $k_x$-$k_z$ sampled, with the red arc as seen by the detector, and the $k$-conservation in a direct transition being indicated. **(b)** A $k_x$-$k_y$ ARPES map over energy interval A in Fig. 4(b) encompassing feature 1 near $E_F$. **(c)** A partial representation of the 3D ARPES dataset in $k_x$, $k_y$ and BE. **(d)** A cross section of the data of (b) in $k_y$-BE. **(e)** A cross section of the data in (b) in $k_x$-BE. Experimental data are from the SLS.

**Figure 8 (color; 2-column)**

hν = 833 eV

Figure 8. As Figures 7(c)-(e), but for a non-resonant excitation energy of 833 eV, and with a larger spectrometer acceptance angle that leads to sampling a greater number of BZs. Experimental data are from the ALS.

**Figure 9 (color; 1-column)**

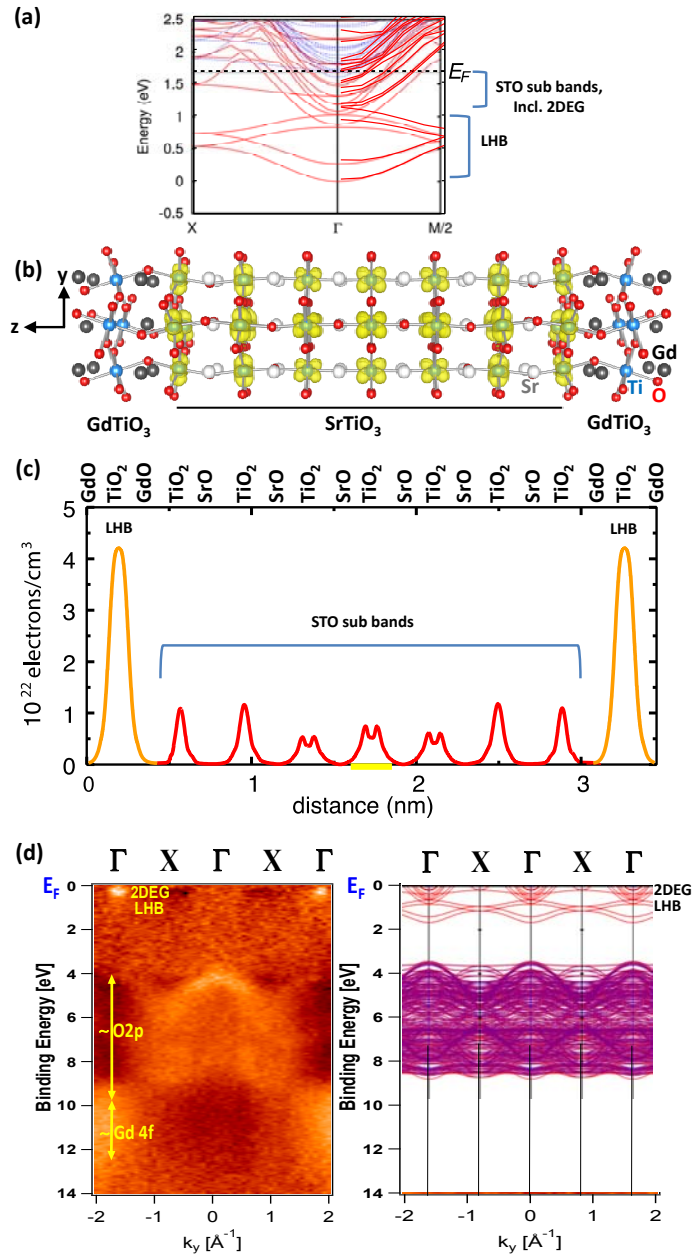

**Figure 9.** Theoretical calculations for the electronic structure of the multilayer sample, including a comparison to experiment. **(a)** Energy bands along the X-Γ-M/2 directions, with the STO subbands, including the 2DEG and the LHB, indicated. **(b)** Charge densities of the occupied electronic states, as isosurfaces plotted at 10% of maximum value: STO conduction bands in yellow, oxygen atoms in red, strontium in white, titanium in blue and gadolinium in black. **(c)** Depth-dependent in-plane averaged charge densities. The central TiO2 layer is underlined in yellow, as in Fig. 1(b). The orange curves correspond to the LHB energy range (in GTO), and the red curves to the 2DEG (in STO). The 2DEG charge density integrates out to the expected ½ electron per unit cell area per each of the two interfaces, or an electron density of 3x10$^{14}$cm$^{-2}$. **(d)** Comparison of a nonresonant kx-BE ARPES map at 1181 eV with calculated bands along Γ-X-Γ-X-Γ, and over a more extended BE range than in (a).

**Figure 10 (color; 1-column)**

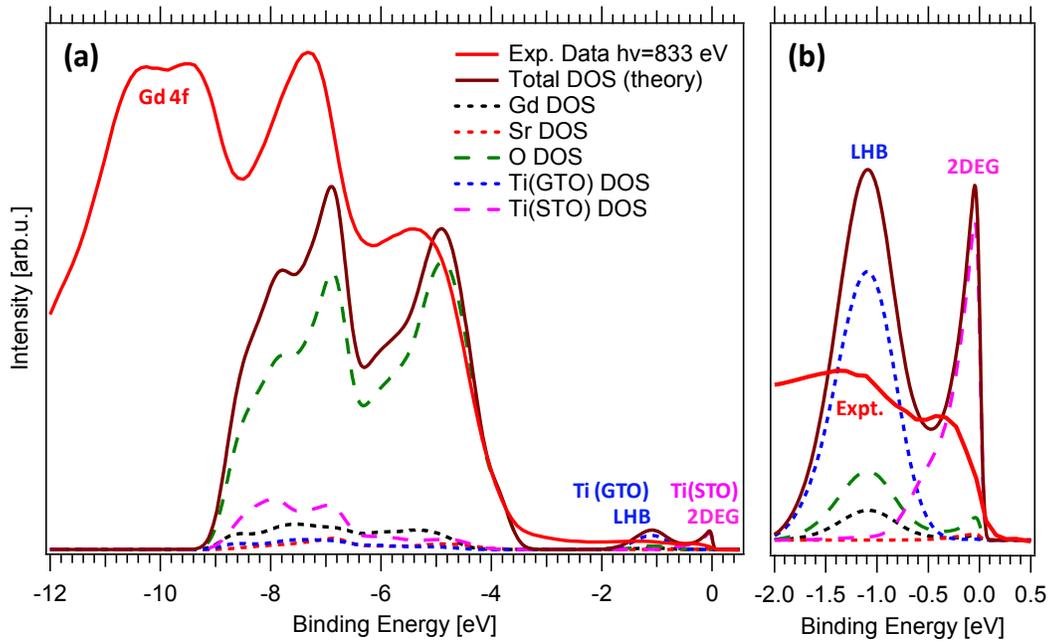

Figure 10.  (a) The experimental MEWDOS VB spectra from the multilayer at 833 eV (Fig. 2(a),(b)) are compared to a simulated spectrum constructed by summing orbital-projected densities of states for the multilayer that are each multiplied by the appropriate differential photoelectric cross section. The Gd 4f states were not included in the calculation.  (b) Enlargement of the region near $E_F$.